\documentclass[fleqn,usenatbib,useAMS]{mnras}
\usepackage{float}
\usepackage{graphicx}	% Including figure files
\usepackage{amssymb}
\usepackage{mathptmx}
\usepackage{gensymb}
\usepackage{color}
\usepackage{epsfig}
\usepackage{mathrsfs}
\usepackage[utf8]{inputenc}

\usepackage[switch]{lineno}
\usepackage[labelsep=endash]{caption}

\def\msun{\,{\rm M}_\odot}

\def\mpc{\,{\rm Mpc}}

\title[]{The Boundary of Cosmic Filaments}

\author[Wang et al.]{
Wei~Wang$^{1,2,3}$, 
Peng~Wang$^{3}$\thanks{E-mail: pwang@shao.ac.cn}, 
Hong~Guo$^{3}$\thanks{E-mail: guohong@shao.ac.cn},  
Xi~Kang$^{4,1}$, 
Noam~I.~Libeskind$^{5}$, 
Daniela Gal\'arraga-Espinosa$^6$,
\newauthor
Volker Springel$^6$,
Rahul Kannan$^{7}$,
Lars Hernquist$^8$,
R\"udiger Pakmor$^6$,
Hao-Ran Yu$^9$,
Sownak Bose$^{10}$,
\newauthor
Quan Guo$^{3}$,
Luo Yu$^{11,1}$,
C\'esar Hern\'andez-Aguayo$^{6,12}$
\\
$^{1}$ Purple Mountain Observatory, Chinese Academy of Sciences, No.10 Yuan Hua Road, 210034 Nanjing, People’s Republic of China.\\
$^{2}$ School of Astronomy and Space Science, University of Science and Technology of China, Hefei 230026, Anhui, People’s Republic of China.\\
$^{3}$ Shanghai Astronomical Observatory, Chinese Academy of Sciences, Nandan Road 80, Shanghai 200030, People’s Republic of China.\\
$^{4}$ Institute for Astronomy, the School of Physics, Zhejiang University, 38 Zheda Road, Hangzhou 310027, People’s Republic of China.\\
$^{5}$ Leibniz-Institut f\"ur Astrophysik Potsdam, An der Sternwarte 16, D-14482 Potsdam, Germany.\\
$^{6}$ Max-Planck-Institut f\"ur Astrophysik, Karl-Schwarzschild-Str. 1, D-85748 Garching, Germany.\\
$^{7}$ Department of Physics and Astronomy, York University, 4700 Keele Street, Toronto, ON M3J 1P3, Canada.\\
$^{8}$ Harvard-Smithsonian Center for Astrophysics, 60 Garden St, Cambridge, MA 02138, USA.\\
$^{9}$ Department of Astronomy, Xiamen University, Xiamen, Fujian 361005, People’s Republic of China.\\
$^{10}$ Institute for Computational Cosmology, Department of Physics, Durham University, South Road, Durham, DH1 3LE, UK.\\
$^{11}$ Department of Physics, School of Physics and Electronics, Hunan Normal University, Changsha 410081, People’s Republic of China. \\
$^{12}$ Excellence Cluster ORIGINS, Boltzmannstrasse 2, D-85748 Garching, Germany.\\
}

\date{Accepted XXX. Received YYY; in original form ZZZ}

\pubyear{2024}

\begin{document}
% Don't change these lines

\label{firstpage}
\pagerange{\pageref{firstpage}--\pageref{lastpage}}

\maketitle

%--------------------------
%  main body
%--------------------------
\begin{abstract}  %
For decades, the boundary of cosmic filaments has been a subject of debate. In this work, we determine the physically motivated radii of filaments by constructing stacked galaxy number density profiles around the filament spines.  We find that the slope of the profile changes with distance to the filament spine, reaching its minimum at approximately $1\mpc$ at $z=0$ in both state-of-the-art hydrodynamical simulations and observational data. This can be taken as the average value of the filament radius. Furthermore, we note that the average filament radius rapidly decreases from $z=4$ to $z=1$, and then slightly increases. Moreover, we find that the radius of the filament depends on the length of the filament, the distance from the connected clusters, and the masses of the clusters. These results suggest a two-phase formation scenario of cosmic filaments. The filaments experienced rapid contraction before $z=1$, but their density distribution has remained roughly stable since then. The subsequent mass transport along the filaments to the connected clusters is likely to have contributed to the formation of the clusters themselves.
\end{abstract}

\section{Introduction}

The origin and evolution of cosmic structures are central issues in the $\Lambda$CDM cosmological model. According to standard structure formation theory, the structures we observe today began as minor fluctuations in the early universe that were subsequently amplified by gravitational instability. Observations \citep{1986ApJ...302L...1D, 2000AJ....120.1579Y, 2001MNRAS.328.1039C} and simulations \citep{2005Natur.435..629S,2020NatRP...2...42V, 2015A&C....13...12N, 2018MNRAS.475..676S, 2019ComAC...6....2N} have revealed the presence of a web-like arrangement of matter in the current universe, known as the cosmic web \citep{1996Natur.380..603B}. This web can be thought of as being composed of four components: clusters, filaments, walls, and voids.

A significant portion of cosmic matter is found in the filaments \citep{2010ApJ...723..364A, 2010MNRAS.408.2163A,  2014MNRAS.441.2923C}, and there is a growing consensus that they play a major role in the formation and evolution of galaxies. This influence is evident in various properties of galaxy, including their mass, shape \citep{2017A&A...600L...6K}, star formation rate \citep{2017A&A...600L...6K}, spatial alignment \citep{1993ApJ...418..544V, 2007ApJ...655L...5A, 2007MNRAS.375..489H, 2009ApJ...706..747Z, 2014MNRAS.443.1090F}, abundance of satellite galaxies \citep{2015ApJ...800..112G}, and correlation of angular momentum \citep{2007ApJ...655L...5A, 2007MNRAS.375..489H, 2012MNRAS.427.3320C, 2013ApJ...762...72T, 2014MNRAS.444.1453D, 2015ApJ...798...17Z, 2017MNRAS.468L.123W, 2018ApJ...859..115W, 2018ApJ...866..138W}. Several different algorithms have been developed to identify filaments based on the distribution of galaxies. However, most algorithms have limitations in extracting spatial information and describing filaments as one-dimensional structures without considering their radial extent \citep{2018MNRAS.473.1195L}. Without a clear understanding of the boundary of filaments, it is difficult to quantify how filaments affect the properties of the galaxy \citep{2017MNRAS.465.3817M, 2018MNRAS.474.5437L, 2018MNRAS.474..547K, 2019A&A...632A..49S}.

The density profiles of mass or galaxy number around the spines of filaments have been investigated in both observations and simulations \citep{2005MNRAS.359..272C, 2006MNRAS.370..656D, 2010ApJ...723..364A, 2010MNRAS.408.2163A, 2010MNRAS.407.1449G, 2010MNRAS.409..156B, 2014MNRAS.441.2923C, 2020A&A...638A..75B, 2020A&A...641A.173G, 2020A&A...637A..41T}. These density profiles can be fitted with various functional forms, such as power or exponential laws, to determine the scale radii that describe the typical sizes of filaments. However, the resulting radii can range from $0.1~\mpc$ to $10~\mpc$, depending on the fitting functions used. On the other hand, observations \citep{2021NatAs...5..839W} and simulations \citep{2021MNRAS.506.1059X} have also observed the spin of filaments, with the rotation curves peaking approximately $1 \mpc$ from the filament spine and then decreasing to zero approximately $2 \mpc$. This suggests that filaments may have a physical radius of around $1 \mpc$ to $2 \mpc$, which is also commonly adopted as the typical size of filaments \citep{2005MNRAS.359..272C}. Having a well-defined physical radius for filaments would then greatly enhance our understanding of their formation and evolution.

\section{Data}

This study is based on observed galaxy samples in the local universe, and these observations are compared to the results of advanced hydrodynamical simulations. The following sections will provide a summary of the data, the sample selection process, the algorithm used to identify filaments, the definition of the filament radius, and the estimation of errors.

\subsection{Observational data}
We adopted two galaxy catalogues obtained from the Sloan Digital Sky Survey \citep[SDSS;][]{2000AJ....120.1579Y}. One of the catalogues is derived from the New York University Value-Added Galaxy Catalogue \citep[NYU-VAGC;][]{2005AJ....129.2562B} of SDSS Data Release 7 \citep{2009ApJS..182..543A}, while the other is obtained from the galaxy group catalogue \citep{2017A&A...602A.100T} constructed using SDSS Data Release 12 \citep{2015ApJS..219...12A}. The SDSS covers a large area that spans more than a quarter of the sky, providing comprehensive imaging, photometric, and spectroscopic data for a dense sample of galaxies. This allows for a statistically robust analysis of cosmic filaments. Both catalogues comprise approximately 600,000 galaxies with spectroscopic redshifts ranging from 0.01 to 0.2. In the NYU-VAGC catalogue, the galaxy redshifts are not corrected for redshift space distortion (RSD), whereas in the galaxy group catalogue, this correction is properly accounted for. For the RSD correction, each galaxy in the group catalogue is first assigned to a group identified with a modified Friends-of-Friends halo finder \citep[FoF;][]{1982ApJ...257..423H, 1985ApJ...292..371D, 2001MNRAS.328..726S, 2009MNRAS.399..497D}, and then the galaxy positions along the line-of-sight are corrected using the radial velocity dispersion of the group and the group size on the sky plane \citep[see details in Appendix C of][]{2014A&A...566A...1T}. It is demonstrated in \cite{2017A&A...602A.100T} that the RSD effect is reasonably removed with such a procedure.

\subsection{Numerical simulations}
For our numerical simulation sample, we used the galaxy catalogue of the MillenniumTNG project (MTNG; \citep{2023MNRAS.524.2556H,2023MNRAS.524.2594K,2023MNRAS.524.2539P}). The cosmological parameters adopted by MTNG were taken from the Planck Collaboration \citep{2016A&A...594A..13P}, which included $\Omega_{m}=0.3089$, $\Omega_{\Lambda}=0.6911$, $\Omega_{b}=0.0486$, and $h=0.6774$. We used the hydrodynamical run with the largest box size, MTNG740, which has a box size of $\rm (500 \ cMpc/h)^3$ or $(\rm \ 738.1 cMpc)^3$, identical to the Millennium simulation \citep{2005Natur.435..629S}, apart from differences in the cosmological parameters. The MTNG run contained $4320^3$ dark matter particles and $4320^3$ initial gas cells, with mass resolutions of $1.7\times 10^8\,{\rm M_{\odot}}$ and $3.1\times 10^7\,{\rm M_{\odot}}$, respectively.  The Friends-of-Friends algorithm \citep[FoF;][]{1982ApJ...257..423H, 1985ApJ...292..371D, 2001MNRAS.328..726S, 2009MNRAS.399..497D} was first applied to dark matter particles, and then the corresponding baryonic matter was assigned to the same groups as the closest dark matter particle, followed by an application of the SUBFIND-HBT \citep{2021MNRAS.506.2871S} substructure finder. The galaxy positions were determined by the most bound particle of each subhalo.

\subsection{Sample selection}
Taking into account the flux limits of SDSS galaxy samples \citep{2004MNRAS.351.1151B, 2011MNRAS.418.1587T} and the mass resolution of the MTNG simulation, we only selected galaxies with a stellar mass greater than $10^{9}\rm M_\odot$ for this study. 
To investigate the effect of RSD, we selected the same set of galaxies in the two SDSS catalogues with and without RSD corrections. We focus only on galaxies more massive than $10^9\msun$ in the redshift range of $0.01<z<0.1$. The final two SDSS catalogues, labelled as \texttt{SDSS(non-RSD)} and \texttt{SDSS(RSD)}, contain 261,354 galaxies.
To further account for the incompleteness of low-mass SDSS galaxies, we calculate the galaxy number density profile in SDSS by weighting each galaxy by $1/V_{\rm max}$, where $V_{\rm max}$ is the maximum accessible volume for each galaxy as determined in the NYU-VAGC catalogue \citep{2005AJ....129.2562B}. 

At redshift $z=0$, the MTNG contains a total of 3,083,441 galaxies with a stellar mass greater than $10^9\msun$. To account for the RSD effect, we used a plane-parallel approximation with the line of sight (LOS) along the $\hat{z}$ direction. The positions of galaxies along the $\hat{z}$ direction were further distorted by their corresponding peculiar velocity, $v_{\hat{z}}$. The MTNG samples with and without the RSD effect are denoted as \texttt{MTNG(non-RSD)} and \texttt{MTNG(RSD)}, respectively. To consider the effect of different mass thresholds , we adopt the mass range from $10^{8.5}\msun$ to $10^{10}\msun$ at z=0 and z=4 in MTNG and z=0 in SDSS. For details, we refer readers to the Figure~\ref{fig:f1_method} in Section~\ref{sum_dis}.

\section{Method}
\subsection{Filament finder}
We used the Discrete Persistent Structures Extractor \citep[hereafter \texttt{DisPerSE;}][]{2011MNRAS.414..350S,2011MNRAS.414..384S} -- one of the most widely used cosmic web finders -- to identify cosmic filaments in this study. To extract the filaments using DisPerSE, we first estimate the underlying density field using the Delaunay tessellation field estimator \citep[hereafter DTFE,][]{2000A&A...363L..29S, 2009LNP...665..291V} by tracing the density field with a galaxy distribution instead of a matter distribution, due to the observation-driven approach. The DTFE densities are then smoothed once by the average density of the surrounding vertices of a given vertex by applying the \texttt{netconv} function to reduce the contamination of the shot noise. Lastly, filaments are identified using the \texttt{mse} function in the DTFE density field. The \texttt{mse} function identifies the critical points where the density gradient vanishes. Filaments are sets of segments connecting maximum-density critical points to saddles along the density field's ridges. The persistence significance level of filaments can be controlled by adjusting the ratio of the density of the two critical points in the pair \citep{2011MNRAS.414..350S,2011MNRAS.414..384S}. 

The effects of different \texttt{DisPerSE} parameters have been carefully investigated in literature \citep[see e.g.,][]{2020A&A...634A..30M, 2020A&A...642A..19M,Galarraga2024}. We follow the fiducial parameters suggested by \cite{Galarraga2024} to extract the filaments. In summary, we apply a one-time smoothing of the DTFE density field and set the persistence significance level at 2$\sigma$. For the sake of fair comparisons among all the data in this paper, we adopt the same parameters to identify filaments in both real and redshift spaces. 
The ``filament'' defined in \texttt{DisPerSE} connects the maximum point (density peak) or bifurcation point (the cross point of two filaments) and the saddle point of the density along the direction of the density gradient. We follow the strategy of \cite{2020A&A...634A..30M} and treat the bifurcation points as nodes, since they can be considered as unresolved clusters. After the extraction, we apply a final smoothing for the extracted skeleton to eliminate the effect of shot noise \citep{2020A&A...634A..30M, 2020A&A...642A..19M}. 
We then combine two such filaments that share the same saddle point into a single one to ensure that the filaments connect the nodes. We refer the readers to \cite{2011MNRAS.414..350S} for more details. 
With the lower stellar mass threshold of $10^9\msun$, we found 9,839 and 5,758 filaments in the \texttt{SDSS(non-RSD)} and \texttt{SDSS(RSD)} samples, respectively. Similarly, in the \texttt{MTNG(non-RSD)} and \texttt{MTNG(RSD)} samples, we found 138,737 and 80,384 filaments, respectively.
 
\subsection{Definition of filament radius}
Analogous to the definition of the splashback radius of a dark matter halo \citep{2014JCAP...11..019A, Diemer2014, 2015ApJ...799..108D, 2015ApJ...810...36M}, we can determine the physical filament radius from positions with the lowest logarithmic gradient in the galaxy number density profile around the filament spines. As in previous studies \citep{2020A&A...641A.173G, Galarraga2024, 2022MNRAS.516.6041Y}, we construct the number density profiles of galaxies around filament spines by counting galaxies in hollow cylindrical shells of increasing radii. The average galaxy number density profiles are then obtained by stacking the measurements for all filament segments. To ensure that the resulting galaxy number density profile is not influenced by nodes, we exclude filament segments that are within 2 Mpc of any nodes (i.e. the distance to the end point of filament) identified in \texttt{DisPerSE}. Although the galaxy density profiles around the filaments would be affected by the different choices of \texttt{DisPerSE} parameters and the stacking method, we verify that our derived filament radius is not significantly affected, which makes it robust to the adopted parameters. Detailed tests are presented in Appendix~\ref{app:test}.

\subsection{Error estimation}
Throughout this paper, we estimate the errors of the galaxy number density profiles using the jackknife resampling technique in both simulation and observational data \citep{2015MNRAS.453.4368G}. In detail, we divide the SDSS galaxies into 32 subvolumes with the same sky area and divide the MTNG galaxies into 32 subboxes of equal volume. The error $\sigma_\rho$ on the galaxy density profile $\rho$ can then be estimated as follows,
\begin{equation}
    \sigma_\rho^2=\frac{N-1}{N}\sum_{i=1}^{N}(\rho_i-\bar{\rho})^2
\end{equation}
where $N=32$ is the number of jackknife subsamples and $\rho_i$ is the galaxy density profile measured for the $i$-th subsample.

\section{Analysis \& Results}
%%%%\section*{Results}
%-------------------------
% Results
%------------------------
\begin{figure*}
\centerline{\includegraphics[width=0.4\textwidth]{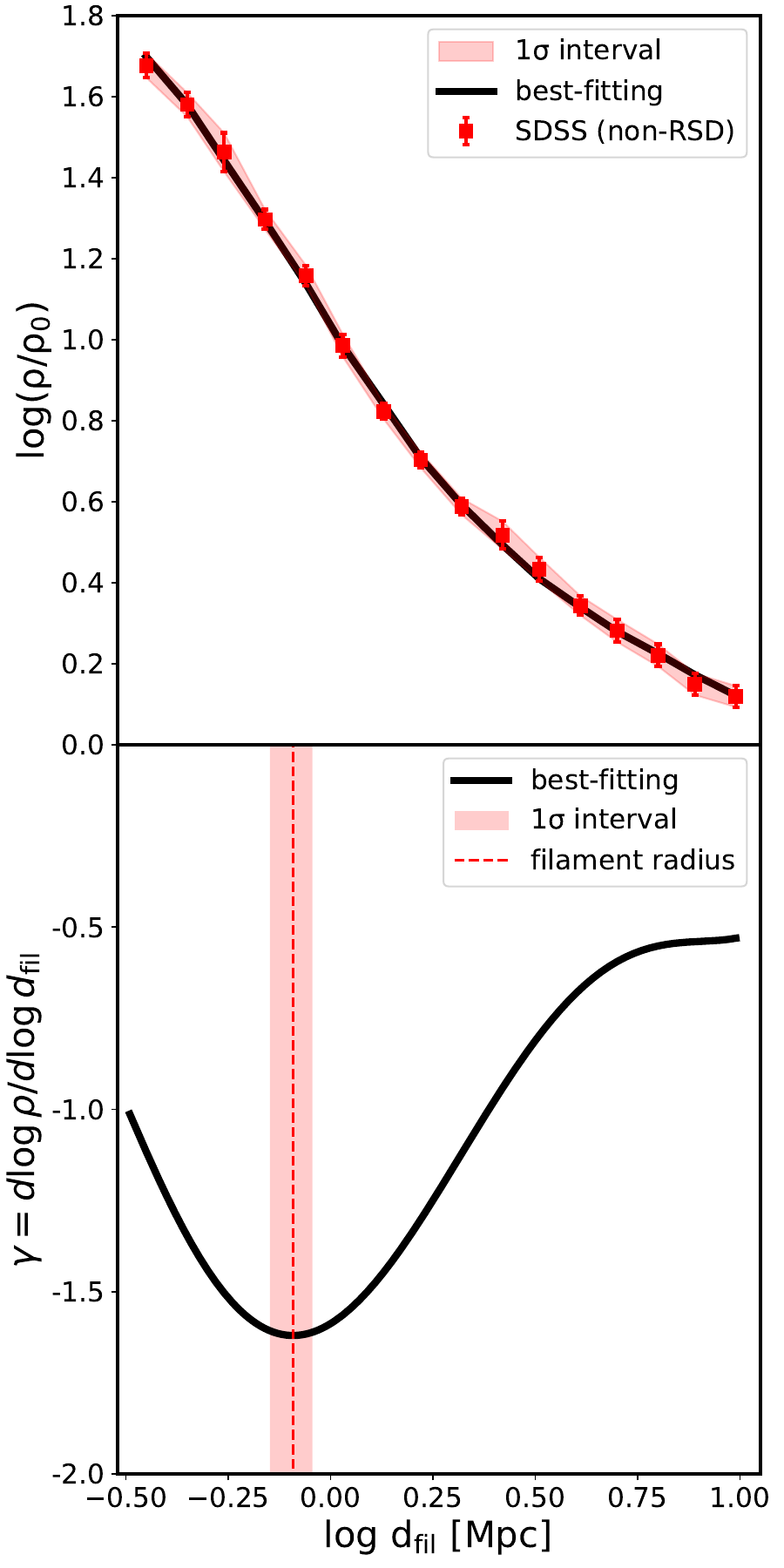}\includegraphics[width=0.4\textwidth]{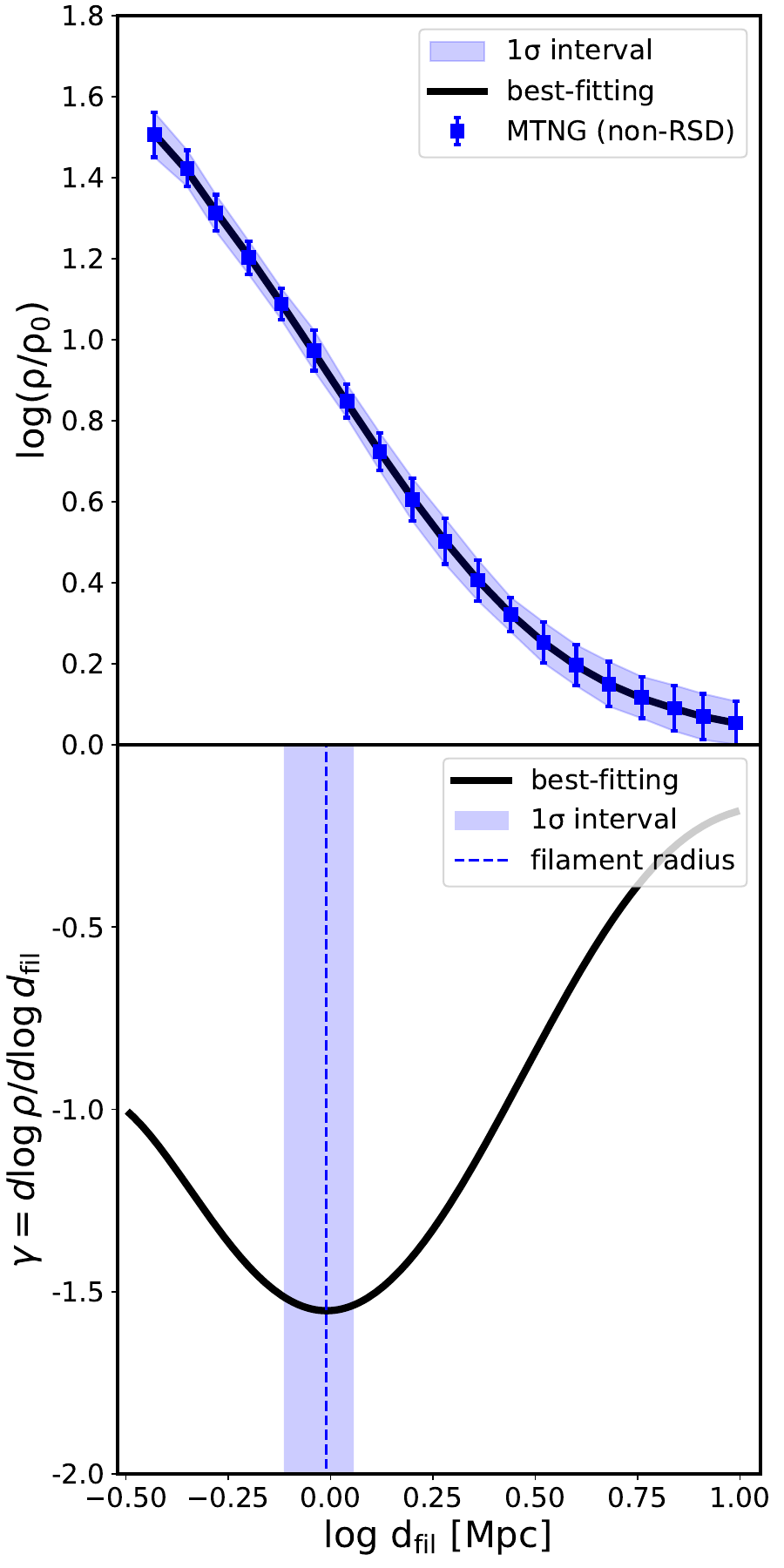}}
\caption{\textbf{Upper panels}: the relative number density profile of galaxies around their host cosmic filaments, $\rm \rho/\rho_0$, as a function of the perpendicular distance to the filament spine, $d_{\rm fil}$, where $\rho_0$ is the background density of galaxies above a threshold mass of $10^9\msun$ in the sample. The colour symbols with error bars are measured from SDSS (left) at $0<z<0.1$ and MTNG (right) at $z=0$, respectively. The errors are measured from the Jackknife resampling method of 32 subsamples with equal volumes. Solid black lines show the best-fitting models using the MCMC method. \textbf{Bottom panels}: The logarithmic slope of the number density profile, $\gamma\equiv d\log\rho/d\log d_{\rm fil}$, derived directly from the best-fit profile shown in the upper panels. The mean values of the filament radius are represented by vertical dashed lines ($0.81\mpc$ for SDSS and $0.98\mpc$ for MTNG), while the shaded area illustrates the 16th to 84th percentile range obtained from the MCMC chains.}
\label{fig:f1}
\end{figure*}

This paper investigates cosmic filaments, which can be thought of as thin and elongated cylinders or curvilinear structures, identified from the galaxy distribution using the popular filament finder \texttt{DisPerSE} \citep{2011MNRAS.414..350S, 2011MNRAS.414..384S}. In essence, the \texttt{DisPerSE} approach identifies ``critical points'' (in the Morse theory sense) as locations where the gradient of the density field vanishes, and connects these to identify ridges or filaments. The framework allows the number of galaxies in each cylinder shell to be counted as a function of the distance from the filament spine, known as the galaxy number density profile \citep{2020A&A...641A.173G, Galarraga2024, 2022MNRAS.516.6041Y}. Since the absolute number of galaxies depends on the redshift and simulation resolution, the number density relative to the background number density provides a more meaningful representation of the local environment. The background number density is obtained simply by dividing the total number of galaxies by the volume. 

%In this study, we focus on the cosmic filaments identified in the state-of-the-art hydrodynamical simulation MillenniumTNG (hereafter MTNG) with a volume of $\sim740^3\,{\rm Mpc}^3$ (see Methods for details) \citep{2023MNRAS.524.2556H, 2023MNRAS.524.2539P}. 
This study focuses on two sets of galaxy catalogues, one obtained from observation and the other from simulation. The analysis of cosmic filaments begins with a publicly available galaxy group catalogue \citep{2017A&A...602A.100T} derived from the Sloan Digital Sky Survey (SDSS) Data Release 12 \citep{2015ApJS..219...12A}. It is worth mentioning that the redshift space distortion (RSD) effect in this catalogue has been reasonably corrected, enabling an easier comparison with the simulation results in real space. The theoretical predictions are then examined using the MTNG with a volume of approximately $740^3\,{\rm Mpc}^3$. The evolution of the filament radius is investigated in this simulation. All the filament radii in the following sections are in comoving units.

\subsection{Filament Radius at $z=0$}
The upper panels of Figure~\ref{fig:f1} show the average radial profiles of the galaxy number density distribution around the filaments, $\rho(d_{\rm fil})/\rho_0$,  for both SDSS (left panel, $0<z<0.1$) and MTNG (right panel, $z=0$) using galaxies with stellar masses of $M_\ast>10^9\msun$ (see Figure~\ref{fig:f1_method} for the dependence on the stellar mass threshold). $\rho_0$ is the background galaxy number density above the stellar mass cutoff and $d_{\rm fil}$ is the perpendicular distance from the filament spines. It is evident that the number density of galaxies close to filaments is much higher than the average cosmic density, with the central region of the filament having a density more than 50 times higher than the background. As one moves away from the spine of the filaments, the density gradually decreases, although the rate of decrease is not uniform at different distances, in agreement with previous studies also using the TNG set of simulations \citep{2020A&A...641A.173G, Galarraga2024}. 

In principle, the logarithmic slope of the number density profile, $\gamma\equiv d\log\rho/d\log d_{\rm fil}$, depends on the concentration of matter within the filaments. The variation of $\gamma$ with $d_{\rm fil}$ indicates how fast the galaxy number density would decrease away from the filament spines. This is analogous to the slope of the matter density profile of dark matter haloes. The radius corresponding to the minimum of the dark matter radial density gradient is proposed to be a physically motivated halo boundary \citep{Diemer2014,2015ApJ...799..108D,2015ApJ...810...36M}, known as the splashback radius where dark matter particles reach the apocenters of their first orbits. Similarly, we can essentially use the slope $\gamma$ to define the average filament radius, where the value of $d_{\rm fil}$ with the minimum $\gamma$ can be seen as the effective radius of the filaments. However, we caution that the structure of filament is not fully virialized as in the case of dark matter halos. We will explore the physical meaning of such a definition in our future work.

To accurately determine the logarithmic slope $\gamma$, we fit a sixth-order polynomial function to $\log (\rho/\rho_0)$, which is represented by the solid black line in the upper panel of Figure~\ref{fig:f1}. We find that the choice of sixth-order polynomials efficiently reduces the noise in the profile of $\rho(d_{\rm fil})$ without affecting the actual values of $\gamma$ (see detailed tests in Appendix~\ref{app:para}), similar to the method used in previous studies focusing on haloes \citep{Diemer2014}. We then apply the Markov Chain Monte Carlo (MCMC) method to explore the parameter space. The slope $\gamma(d_{\rm fil})$ is then derived from the best-fitting function, as shown in the bottom panels of Figure~\ref{fig:f1}. We observe that $\gamma$ reaches its minimum at approximately $d_{\rm fil}\sim1\,{\rm Mpc}$, the scale that we define as the average radius of the filament ($R_{\rm fil}$). The scatter of the filament radius is determined from the 16th and 84th percentile ranges of the MCMC chains. This value of $R_{\rm fil}$ is similar to the common rough estimate of filament sizes in the literature \citep{2005MNRAS.359..272C, 2010ApJ...723..364A, 2010MNRAS.409..156B, 2010MNRAS.407.1449G, 2014MNRAS.441.2923C}, but here we derive it precisely from the number density distribution profiles. We emphasise that the minimal $\gamma$ is around $-1.7$, which is not as steep as the halo density profile at the splashback radius (around $-3$). This implies that the edges of the filaments are not as sharp as those of haloes.

\subsection{Effect of Redshift Space Distortion}
The spatial distribution of galaxies can be influenced by RSDs, which alter the galaxy number density profile around filaments. To quantify this effect, we also measure $R_{\rm fil}$ with the RSD-included $\rho(d_{\rm fil})$ in MTNG. We assumed a plane-parallel model with the $\hat{z}$ direction as the line of sight (LOS) and we changed the galaxy positions along the $\hat{z}$ direction using the peculiar velocity of the LOS $v_{\hat{z}}$. The results are shown in Figure~\ref{fig:f2}. The galaxy density profiles and the corresponding slopes are shown in the Appendix~\ref{app:rsd}. Additionally, we determined the filament radius in the same manner for the observed galaxy samples of SDSS, but without applying the RSD correction. This allows us to compare the impact of RSD on the determination of the filament radius.

\begin{figure}
\centerline{\includegraphics[width=0.45\textwidth]{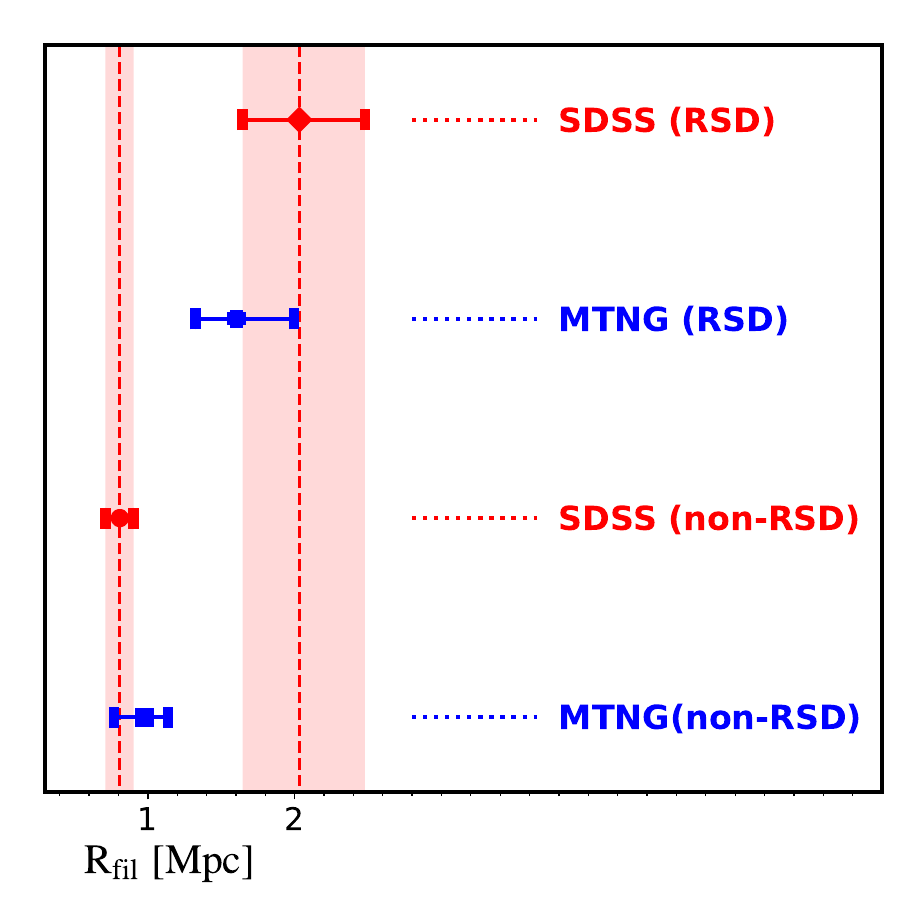}}
\caption{\textbf{Comparison of the radius of the filament in several different sets of observed and simulated data.} Observational data is represented by red symbols, while data from the MTNG simulation is represented by blue symbols. Both the observational and simulation data take into account the effect of Redshift Space Distortions (RSD). The filament radius in SDSS is indicated by two red vertical lines, and the red bars (which are the same as the error bars) represent the $1\sigma$ scatter. The scatter was estimated from the MCMC chains of the fittings to the galaxy number density profiles. In the MTNG simulation, the RSD effect was accounted for by assuming a plane-parallel model with the line of sight (LOS) along the $\hat{z}$ direction. The galaxy positions were changed along the $\hat{z}$ direction using the peculiar velocity of the LOS, $v_{\hat{z}}$. }
\label{fig:f2}
\end{figure}

The measurements of $R_{\rm fil}$ in real and redshift spaces in MTNG are consistent with the observations from SDSS, suggesting that the MTNG simulation accurately represents the distribution of galaxies around filaments.  Inclusion of RSDs causes the radius of the filament to increase from $1\,{\rm Mpc}$ in the real space to approximately $2\,{\rm Mpc}$ in the redshift space. On smaller scales, the RSDs result in the elongation of galaxy distributions along the LOS, a phenomenon known as the Fingers-of-God effect. As a result, the observed filaments in redshift space also appear thicker. Additionally, the number of filaments identified in SDSS decreases significantly from 9,839 in the non-RSD sample to 5,758 in the RSD sample, representing a decrease of approximately 42\%. Similarly, the corresponding numbers for MTNG are 138,737 (non-RSD) and 80,384 (RSD), respectively. This reduction in the number of filaments occurs primarily in short filaments with lengths of a few Mpc, as the identification of these filaments is strongly influenced by RSD.

\begin{figure*}
\centerline{\includegraphics[width=0.98\textwidth]{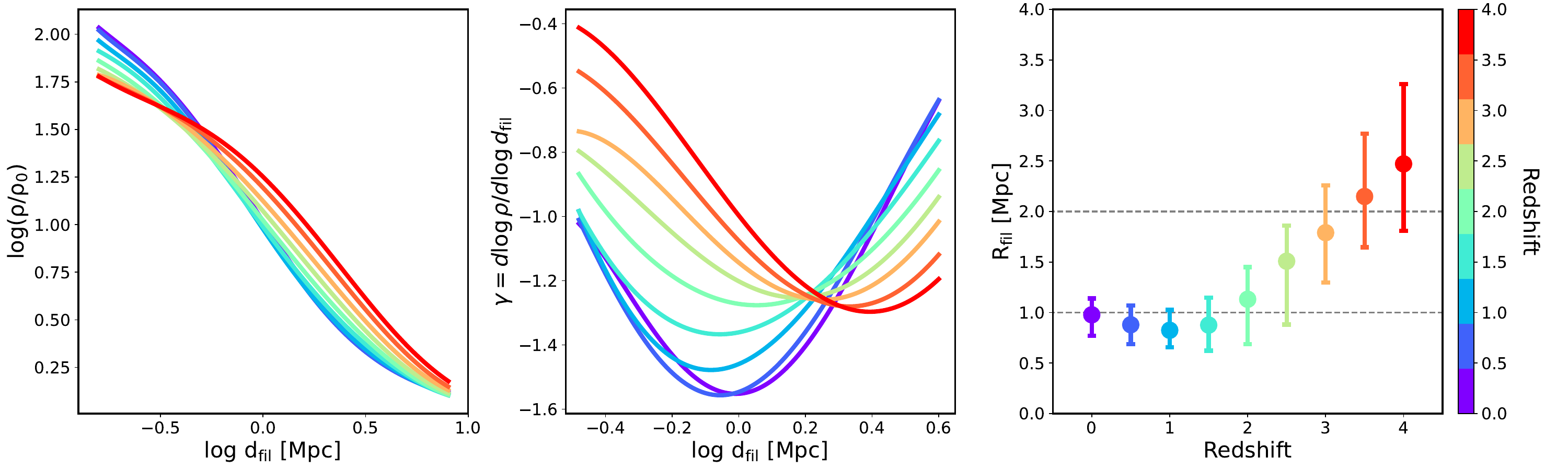}}
\caption{\textbf{Evolution of the galaxy number density profile around filament spines, its slope, and the corresponding filament radius.} Similar to Figure~\ref{fig:f1}, but we consider the redshift dependence of the filament radius. Measurements at different redshifts are shown in different colours, as indicated by the colour bar on the right. \textbf{Left panel:} the density profiles of the galaxy number around filaments as a function of the distance to the filament spine $d_{\rm fil}$ at the corresponding redshifts. \textbf{Middle panel:} the variation of the best-fitting slope $\gamma$ with $d_{\rm fil}$ at different redshifts. \textbf{Right panel:} the best-fitting radius of filaments as a function of redshift. The dashed lines indicate the filament radius of 1 $\mpc$ and 2 $\mpc$, respectively. The average filament radius has undergone a rapid decrease from $z=4$ to $z=1$ and a slow growth afterwards. The slope $\gamma$ at $R_{\rm fil}$ also becomes steeper with time.}
\label{fig:f3}
\end{figure*}

\begin{table*}
	\centering
	\caption{The number of galaxies and filaments in given redshifts in Figure~\ref{fig:f3}. }
	\label{tab:z_0_4_Mth9}
	\begin{tabular}{lccccccccc} % four columns, alignment for each
		\hline
		z & 0.0 & 0.5 & 1.0 & 1.5 & 2.0 & 2.5 & 3.0 & 3.5 & 4.0\\
		\hline
		\# of gal & 3,083,441 & 3,095,094 & 2,869,670 & 2,487,973 & 2,056,615 & 1,601,176 & 1,196,343 & 806,903 & 550,774\\
		\# of fil & 138,737 & 146,513 & 136,752 & 115,764 & 92,817 & 69,943 & 51,230 & 35,074 & 23,794\\
		\hline
	\end{tabular}
\end{table*}

\subsection{Evolution of Filament Radius}
Although we lack observational data in a broad redshift range to quantify the evolution of the filament boundary, we can investigate the trend of $R_{\rm fil}$ with redshift using MTNG. Figure~\ref{fig:f3} displays the redshift evolution of $\rho(d_{\rm fil})$, $\gamma(d_{\rm fil})$ and $R_{\rm fil}$ in MTNG from left to right, respectively. The number of galaxies and filaments in each redshift are listed in Table~\ref{tab:z_0_4_Mth9}. It is evident that filament formation has gone through two stages: a rapid radial collapse before $z=1$ and a slower growth along the radial direction afterward. The median filament radius $R_{\rm fil}$ decreases from approximately 2.5 Mpc at $z=4$ to 0.8 Mpc at $z=1$, representing a 78\% decrease in only 4.4 Gyrs. In the following 7.9 Gyrs from $z=1$ to $z=0$, the median $R_{\rm fil}$ only increases from 0.8 to 1 Mpc. Furthermore, the slope $\gamma$ at $R_{\rm fil}$ decreases from $z=4$ to $z=1$, sharpening the edges of the filaments.

The development of filaments is accompanied by the accumulation of matter towards the filament spines, which is demonstrated by the considerable increase in the number density of galaxies within the filaments as the redshift decreases (left panel of Figure~\ref{fig:f3}), in agreement with \cite{Galarraga2024}. The galaxy number density profile becomes quite consistent since the formation of filaments around $z=1$. This suggests that the structure of cosmic filaments was formed essentially around $z=1$.   

\begin{figure*}
\centerline{\includegraphics[width=0.98\textwidth]{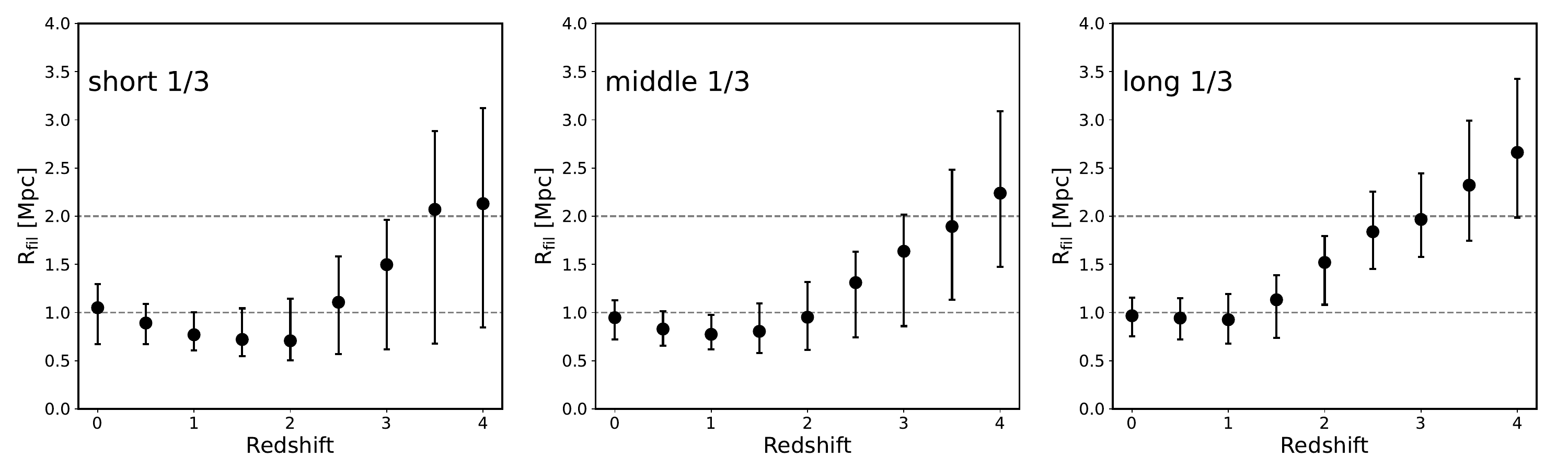}}
\caption{\textbf{Evolution of filament radius with redshift by considering the filament length.} We categorize the entire sample into three sub-samples of equal size based on the length distribution of the filaments. These subsamples are labelled as short, middle, and long filaments, arranged from left to right. At redshift $z=0$, the median filament lengths for the short, middle and long subsamples are 7.56, 14.61, and 25.05 Mpc, respectively. The longer filaments have larger $R_{\rm dfil}$ at $z>1$, and the shorter filaments show a stronger growth of the filament radius.}
\label{fig:f4}
\end{figure*}

\subsection{Dependence on Filament Length}
The length of a filament is an important factor in addition to its radius. It has been demonstrated that shorter filaments tend to have a higher galaxy number density, are located in denser environments, and are connected to more massive objects \citep{2020A&A...641A.173G, 2022A&A...661A.115G}. Therefore, it is crucial to understand how the radius of the filament varies with its length. We divide the filaments into three groups of equal size according to their lengths and present the evolution of their radii in Figure~\ref{fig:f4}. We obtain consistent filament profiles in different lengths with \cite{2020A&A...641A.173G}, i.e. the shorter filaments typically have higher amplitudes of density profiles than longer ones. The evolutionary patterns observed for filaments of different lengths are quite similar, but longer filaments generally have relatively larger $R_{\rm fil}$ (that is, they are thicker) than shorter filaments before $z=1$. Since $z=1$, 
the radius of long filaments is roughly constant (right panel), while the median radii for short filaments are slightly increased (left panel) for $z<1$. This means that the weak growth of the radius of the filament shown in Figure~\ref{fig:f3} is mainly driven by the shortest filaments. This phenomenon is likely attributed to the fact that short filaments are embedded in denser environments \citep{2020A&A...641A.173G, 2022A&A...661A.115G}, facilitating easier matter accretion and growth.  

\begin{figure*}
\centerline{\includegraphics[width=0.45\textwidth]{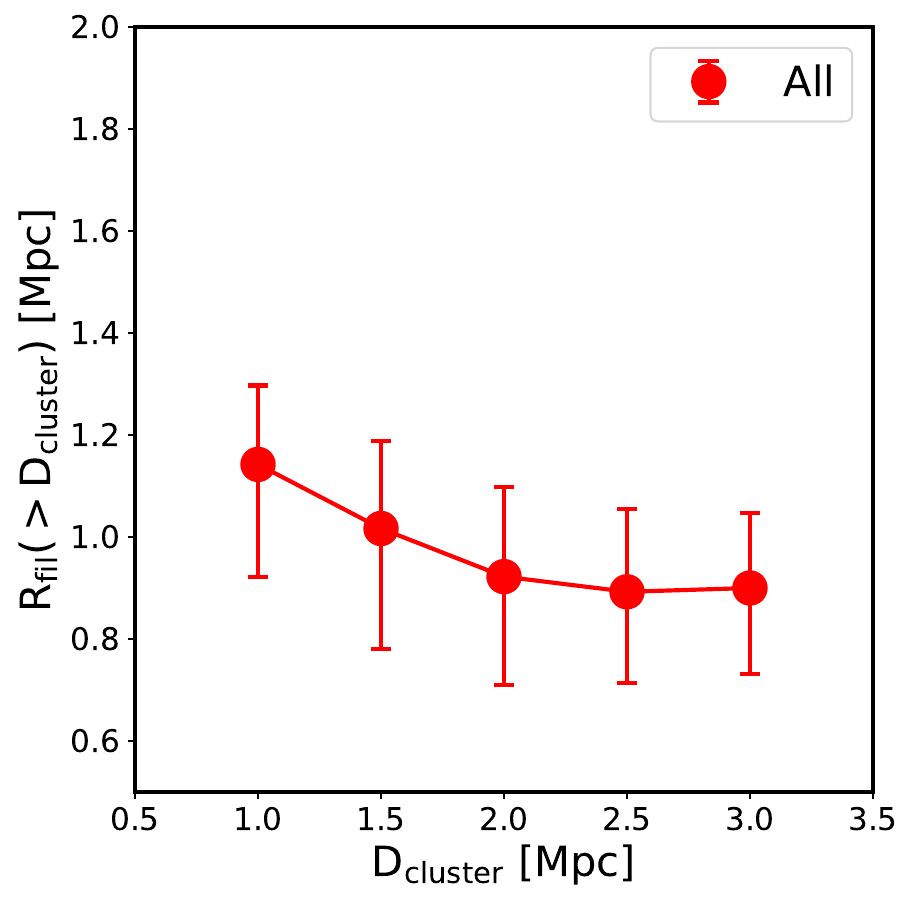}\includegraphics[width=0.45\textwidth]{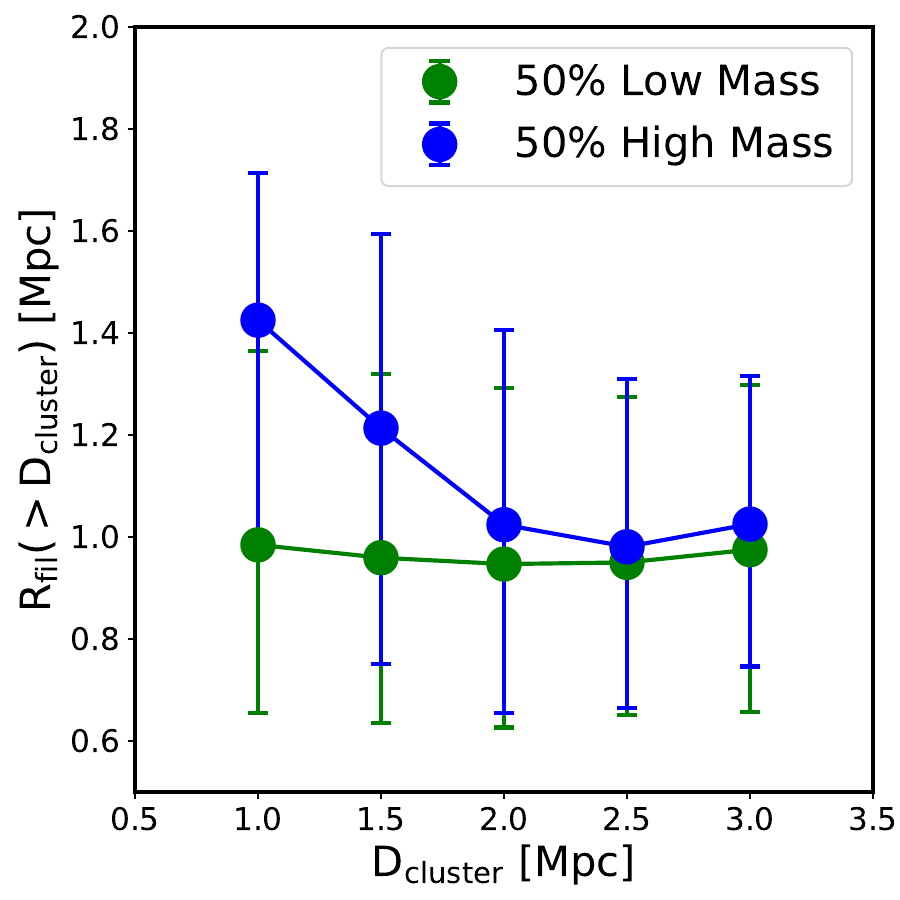}}
\caption{\textbf{The relationship between the filament radius and its distance from the nearest clusters.} \textbf{Left panel:} the filament radius, $R_{\rm fil}$, as a function of the distance, $D_{\rm cluster}$, to the nearest connected cluster. We stacked all filament segments with distances larger than a given threshold of $D_{\rm cluster}$. \textbf{Right panel:} Similar to the left panel, but filaments are subdivided into two subsamples according to the mass of the nearest connected clusters. The blue and green symbols are for the higher-mass and lower-mass halves, respectively.}
\label{fig:f5}
\end{figure*}

\begin{figure*}
%\centerline{\includegraphics[width=\columnwidth]{appendix_f1.pdf}}
\centerline{\includegraphics[width=1.\textwidth]{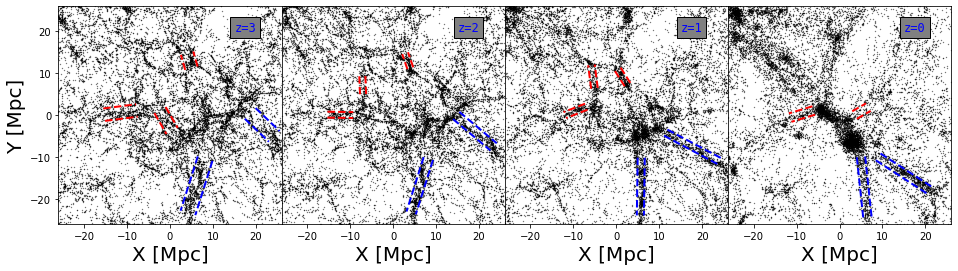}}
\caption{Illustration of the evolution of some selected cosmic filaments around a cluster. The blue and red dashed lines indicate the boundaries of the highlighted long and short filaments, respectively.}
\label{fig:cartoon}
\end{figure*}

This increase in the filament radii of the short filaments is probably due to the accretion of matter near the nodes \citep{1996Natur.380..603B, 2014MNRAS.443.1274L, 2015ApJ...813....6K, 2022ApJ...935..130O}. To investigate this, Figure~\ref{fig:f5} shows the dependence of the filament radius $R_{\rm fil}$ on the distance to the nearest cluster $D_{\rm cluster}$ (left panel) and the further dependence on the cluster mass by dividing the connecting clusters into high and low mass halves (right panel). The identification of clusters at the end points of the filaments is provided by \texttt{ DisPerSE}. We only consider the effect of clusters on the filament radius starting from a distance of $1\mpc$ from the cluster/node centres, which is approximately the halo virial radius of cluster galaxies. We calculate $D_{\rm cluster}$ for each filament segment as the distance from the segment centre to the nearest node. The median radii of the filaments increase toward the clusters, with a growth of around 30\% from a distance of $3\mpc$ to $1\mpc$. The right panel of the figure reveals that the increase in $R_{\rm fil}$ is even more pronounced for more massive clusters, although consistent within the error bars. 

Based on Zel'dovich's theory \citep{1970A&A.....5...84Z} and the dynamics of mass flow within the cosmic web \citep{Icke1991,2014MNRAS.441.2923C}, the cosmic web developed in a hierarchical manner. Initially, matter collapses along the direction of the fastest compression to form extensive cosmic walls. Subsequently, matter flows along the second compression direction, which is within the plane of the wall, to create filaments. The mass flows along the filament direction leading to the final formation of massive clusters. It is crucial to understand that the collapse happens in all three directions at the same time, rather than one after the other, with different collapsing speeds. It is clear from our results that the filaments are basically formed around $z=1$ and the formation of the clusters is supported by the flow of matter along the filaments since $z=1$ \citep{2014MNRAS.441.2923C}. 

In Figure~\ref{fig:cartoon}, we depict the development of a cluster and its neighboring filaments. At $z=3$, the proto-cluster was connected by a few long filaments and many more short ones. We emphasize some filaments with blue (for long filaments) and red (for short filaments) dotted lines to indicate their boundaries. The two long filaments collapsed to form stable structures from $z=3$ to $z=1$, maintaining similar filament radii since $z=1$. The formation of the central cluster accelerated significantly from $z=1$ to $z=0$ after the stable filaments formed, which supports our previous conclusions. The short filaments around the cluster evolved much more dramatically from $z=3$ to $z=0$. Their collapse is still evident from $z=3$ to $z=2$, and the subsequent expansion of their boundaries with the cluster formation is also visible in the figure. Some short filaments merged to become relatively longer filaments, while others merged into the central cluster. Thus, the short filaments at $z=0$ do not exactly correspond to those at $z=3$.

%In contrast to long filaments, short filaments appear to be connected to clusters located in node structures. 
\subsection{Dependence on Stellar Mass Threshold}
\begin{figure*}%[!htp]
%\centerline{\includegraphics[width=\columnwidth]{f6a.pdf}}
\centerline{\includegraphics[width=0.45\textwidth]{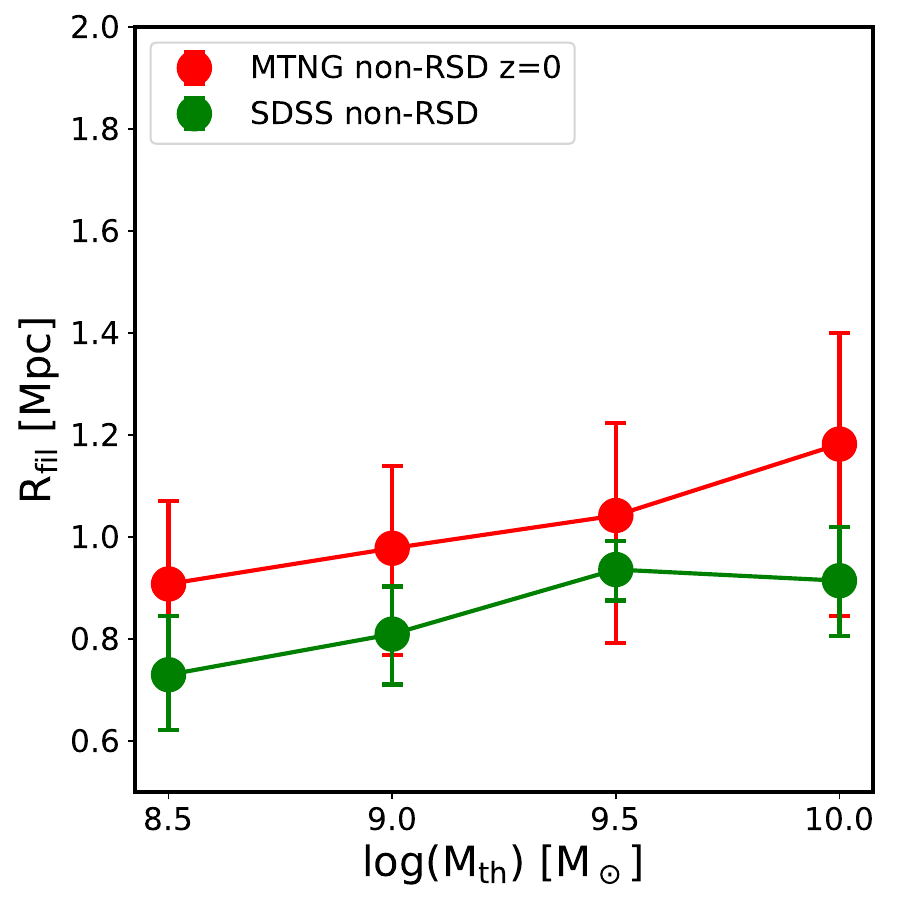}\includegraphics[width=0.45\textwidth]{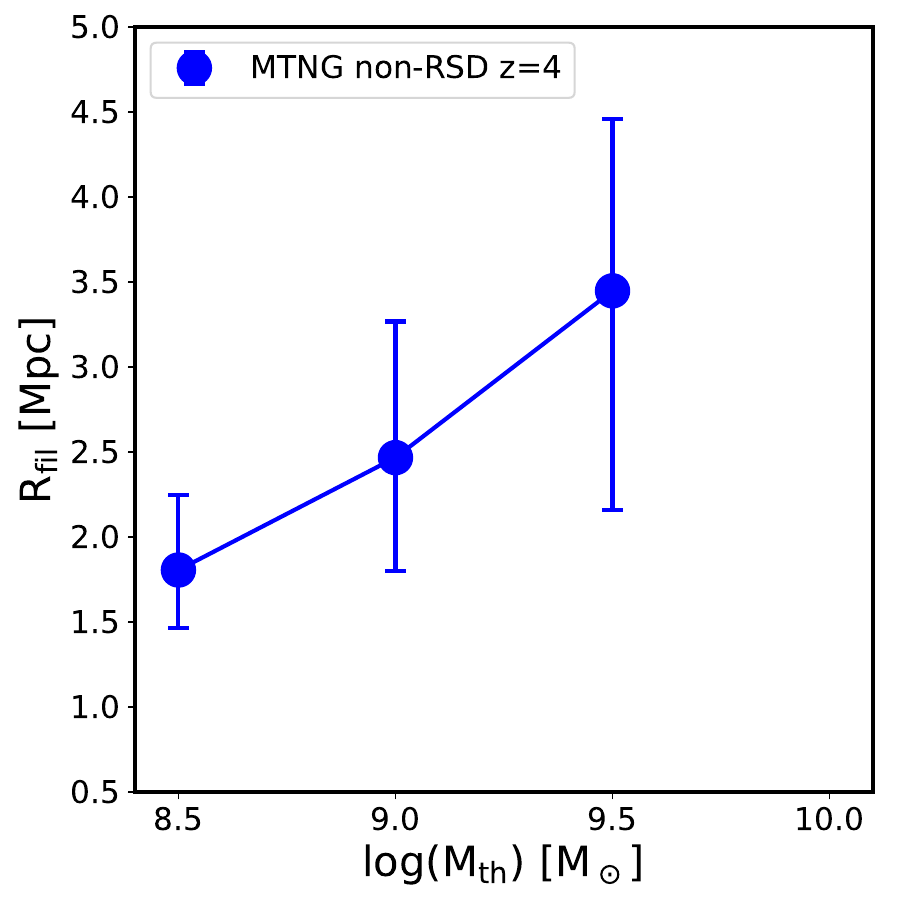}}
\caption{\textbf{Dependence of the average filament radius on the stellar mass threshold.} \textbf{Left Panel:} We measured the average filament radius for the galaxy samples with the lower stellar mass thresholds of $10^{8.5}\msun$, $10^{9.0}\msun$, $10^{9.5}\msun$ and $10^{10}\msun$, for both \texttt{SDSS(non-RSD)} (blue symbols) and \texttt{MTNG(non-RSD)} at $z=0$ (red symbols). \textbf{Right Panel:} The equivalent dependence on the lower stellar mass threshold for \texttt{MTNG(non-RSD)} at $z=4$. A weak increasing trend of filament radius with the stellar mass threshold is found for both observation and simulation at $z=0$, while the dependence at $z=4$ in MTNG is stronger. }
\label{fig:f1_method}
\end{figure*}

In this paper, we adopt the stellar mass threshold of $M\ast>10^9 M_\odot$ to extract the filaments, since SDSS galaxies are roughly complete above this mass threshold. In Figure~\ref{fig:f1_method}, we investigate the relationship between the average radius of filaments and the threshold mass of galaxies for both \texttt{MTNG(non-RSD)} and \texttt{SDSS(non-RSD)}. It is important to note that the threshold mass of galaxies is used to select the sample before conducting the filament search. This results in variations in the filament samples based on different mass thresholds. The average filament radius increases slightly with the threshold mass $M_{\rm th}$. At $z=0$ for \texttt{MTNG(non-RSD)}, the average filament radius ranges from approximately $0.91\,{\rm Mpc}$ for $\rm M_\ast>10^{8.5}\msun$ to $1.18{\rm Mpc}$ for $M_\ast>10^{10}\msun$. \texttt{SDSS(non-RSD)} exhibits a similar increasing trend, with slightly lower average filament radii compared to \texttt{MTNG(non-RSD)}, but still consistent within the large errors. The small offsets between SDSS and MTNG are primarily due to differences in the stellar mass functions of the two samples. We also observed similar trends for the MTNG galaxies at higher redshifts. This consistency aligns with the fact that more massive galaxies tend to inhabit denser environments and reside closer to clusters. Although the level of evolution of the filament radius may differ depending on the tracers used, the general trend remains the same.

On the other hand, as shown in \cite{2018MNRAS.474..547K},  massive galaxies tend to be residing close to filament spines. The filaments identified using these massive galaxies would be tighter, i.e. with smaller filament radii, which seems to be contrary to our conclusions above. However, such discrepancies are simply caused by the sample selection effect. For a given set of filament segments, the average $R_{\rm fil}$ is indeed decreasing with the increasing stellar mass threshold. But in the results of Figure~\ref{fig:f1_method}, we actually include much more filaments with a smaller $M_{\rm th}$. We display the sample sizes of different $M_{\rm th}$ in Table~\ref{tab:Mth_8_5_9_z_0}. The additional filaments introduced in the lower $M_{\rm th}$ samples are typically in lower density environments, leading to a smaller average $R_{\rm fil}$ as suggested by Figure~\ref{fig:f5}.

%\textcolor{red}{, keeping the stellar mass cut of $\rm 10^9 M_\odot$ fixed, which corresponds to the right panel of Figure~\ref{f1_method}. }
\begin{table}
	\centering
	\caption{The number of galaxies and filaments in MTNG (at $z=0$ and $z=4$) and SDSS (with redshift range $0.01<z<0.2$) at different stellar mass cut, namely $\rm 10^{8.5} M_\odot$, $\rm 10^{9.0} M_\odot$, $\rm 10^{9.5} M_\odot$ and $\rm 10^{10.0} M_\odot$.}
	\label{tab:Mth_8_5_9_z_0}
	\begin{tabular}{rcccc} % four columns, alignment for each
		\hline
		$\rm M_{th}$ & 8.5 & 9.0 & 9.5 & 10.0\\
		\hline
		\# of gal (SDSS) & 276,898 & 261,354 & 226,113 & 142,203\\
            \# of gal (MTNG, z=0)  & 4,516,498 & 3,083,441 & 1,940,397 & 1,095,572\\
            \# of gal (MTNG, z=4) & 1,357,155 & 550,774 & 227,059 & - \\
            \# of fil (SDSS) & 10,488 & 9,839 & 8,700 & 5,699\\
		\# of fil (MTNG, z=0) & 190,878 & 138,737 & 92,594 & 54,389\\
            \# of fil (MTNG, z=4) & 58,432 & 23,794 & 9,538 & - \\
		\hline
	\end{tabular}
\end{table}

\section{Summary \& Discussion}\label{sum_dis}
%----------------------------------
% Summary & Discussion
%----------------------------------
In this paper, we have developed a physically-motivated definition of the radius of filaments in terms of the minimum slope of the galaxy number density distribution around the filament spines. This approach with a precise value of the radius allows us to quantify the formation and evolution of filamentary structures. Our analysis reveals that the average radius of cosmic filaments in the MTNG simulation at $z=0$ is approximately $1\mpc$, which agrees with the filament radius derived from SDSS galaxies after correcting for redshift space distortions (RSD) \citep{2015ApJS..219...12A,2017A&A...602A.100T}. The presence of RSDs causes a distortion in the spatial distribution of galaxies, leading to an increase in the filament radius in redshift space to around $2\mpc$, a result that is also supported by the observed SDSS galaxy sample without the RSD correction. Furthermore, we observe a decreasing trend in the slope ($\gamma$) at the radius of the filament from approximately $-1.2$ at $z=4$ to approximately $-1.6$ at $z=0$, indicating sharper filament edges. However, these slopes are still higher than the typical slope at the splashback radius of dark matter halos, suggesting that the boundaries of filaments are less sharp compared to those of halos.

We can observe that filament formation occurs in two distinct phases. Before $z=1$, the filaments undergo rapid collapse, resulting in a significant decrease in their radius from $z=4$ to $z=1$. By $z=1$, the filaments are more collapsed in the radial direction, and the galaxy number density profile remains relatively stable. Subsequently, the continued weak growth of the filament radius is driven by the presence of short filaments that connect to the node structures. Additionally, at $z=0$, filaments that are closer to clusters exhibit larger radii, and this effect seems to be more pronounced for clusters of higher mass. 

We anticipate that the outcome of this study will not depend on the specific hydrodynamical simulation employed to explore the filament radius. The filament characteristics (especially the galaxy number density profiles around the filaments) constructed for the TNG simulation suite, which is very similar to the advanced MTNG simulation used here, were found to be in line with those in other hydrodynamical simulations \citep{2020A&A...641A.173G}. In essence, filament formation is mainly driven by the influence of the gravity of the density field \citep{2020A&A...641A.173G, Galarraga2024, 2022MNRAS.516.6041Y, 2023PhRvD.107b3514S}, which is unlikely to be significantly altered by different implementations of baryonic physics in different simulation models. Although we use the galaxy number density as the tracer of the filament structure in both SDSS and MTNG, it has been shown in the simulations that the dark matter, gas and stellar distribution profiles are very consistent with each other around the scales of filament radius \citep{2022A&A...661A.115G}. With ongoing high-redshift galaxy surveys, we will soon be able to further constrain the evolution of the filament radius using high-redshift observational galaxy samples, which will lend itself to an interesting test of the $\Lambda$CDM simulation models.

\section*{Acknowledgements}
We thank the anonymous reviewer for the constructive comments that significantly improve the presentation of this paper.
This work is supported by the National SKA Program of China (grant No. 2020SKA0110100). PW is sponsored by Shanghai Pujiang Program(No. 22PJ1415100). HG is supported by the CAS Project for Young Scientists in Basic Research (No. YSBR-092) and the science research grants from the China Manned Space Project with NO. CMS-CSST-2021-A02. H.R.Y. is supported by National Science Foundation of China grant No. 12173030. SB is supported by the UK Research and Innovation (UKRI) Future Leaders Fellowship [grant number MR/V023381/1]. CH-A acknowledges support from the Excellence Cluster ORIGINS which is funded by the Deutsche Forschungsgemeinschaft (DFG, German Research Foundation) under Germany's Excellence Strategy -- EXC-2094 --390783311. We acknowledge the use of the High Performance Computing Resource in the Core Facility for Advanced Research Computing at the Shanghai Astronomical Observatory. 

\section*{Data Availability}
The MillenniumTNG simulations will be publicly available on \url{https://www.mtng-project.org} in the future. The data we use in this article will be shared upon reasonable request to the corresponding author. 
The codes used in this study are available from the corresponding authors upon reasonable request.

\bibliographystyle{mnras}
% \bibliography{ref}

\appendix
\section{The choice of order of polynomial}
\label{app:para}
\begin{figure}
\centerline{\includegraphics[width=0.45\textwidth]{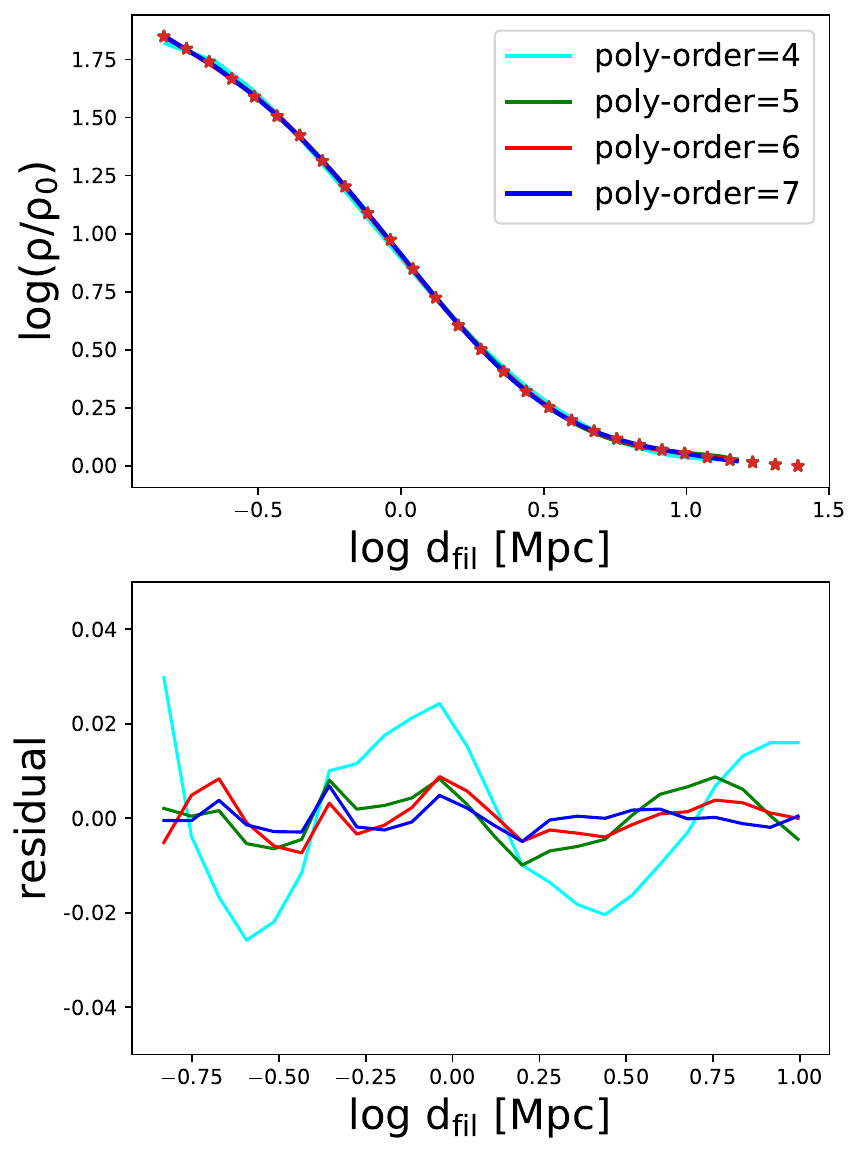}}
\caption{\textbf{Upper panel: }the fittings to the galaxy number density profile of the \texttt{MTNG (non-RSD)} sample using different orders of polynomials, ranging from the fourth-order to the seventh-order. \textbf{Bottom panel: } the percentage residuals between fitting profiles and the data points.}
\label{fig:mq4}
\end{figure}

In this paper, we adopted a sixth-order polynomial function to fit the galaxy number density profiles, which is to minimize the noise in the measurements. In the top panel of Figure~\ref{fig:mq4}, we show the fittings to the galaxy number density profile of the \texttt{MTNG (non-RSD)} sample using different orders of polynomials, ranging from the fourth-order to the seventh-order. The percentage residuals of the fittings are shown in the bottom panel. It is clear that polynomials with more than six orders can fit the profile with sufficient accuracy (better than 1\%).

\section{The profile and slope in redshift space}
\label{app:rsd}
\begin{figure*}
\centerline{\includegraphics[width=0.4\textwidth]{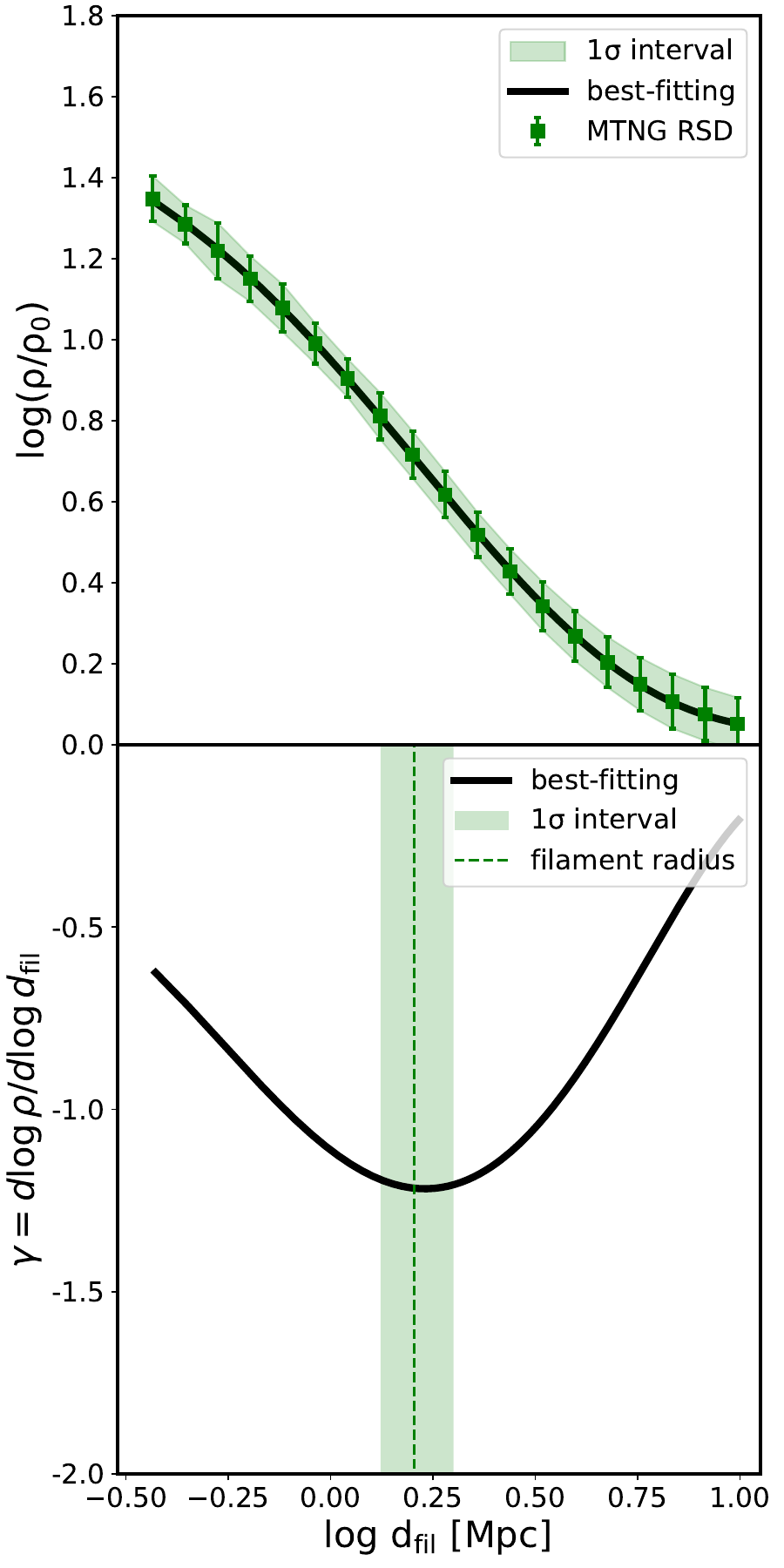}\includegraphics[width=0.4\textwidth]{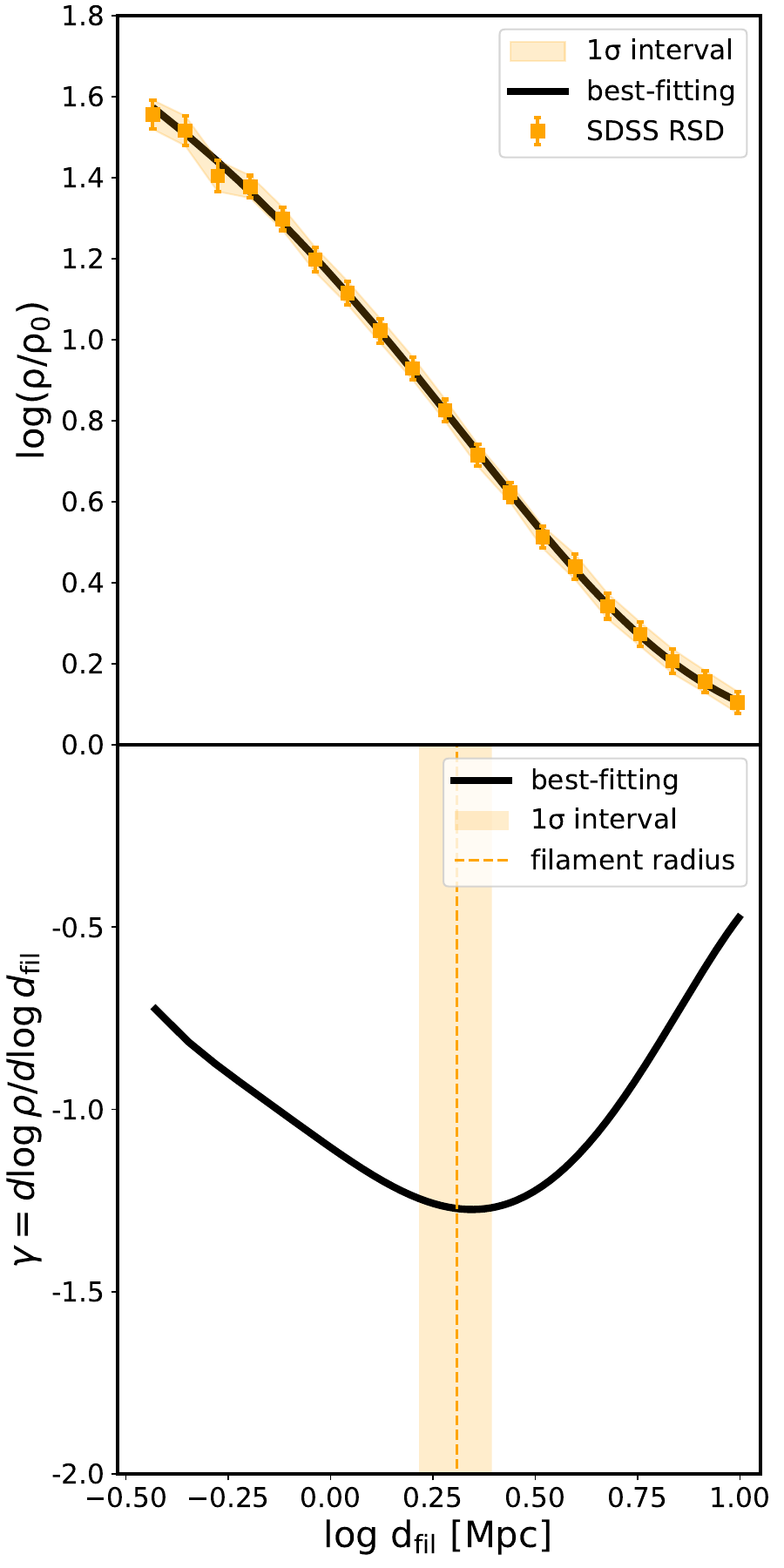}}
\caption{Similar to Figure~\ref{fig:f1} but for the \texttt{MTNG(RSD)} (left panel) and \texttt{SDSS(RSD)} (right panel) samples.}
\label{fig:f1_rsd}
\end{figure*}

Similar to Figure~\ref{fig:f1}, we present the profiles and slopes for the \texttt{MTNG(RSD)} (left panel) and \texttt{SDSS(RSD)} (right panel) samples in Figure~\ref{fig:f1_rsd}. The results of the filament radii are presented in Figure~\ref{fig:f2}. 

\section{Effects of different \texttt{DisPerSE} parameters and profile stacking methods}
\label{app:test}
\begin{figure*}
%\centerline{\includegraphics[width=\columnwidth]{appendix_f1.pdf}}
\centerline{\includegraphics[width=1.\textwidth]{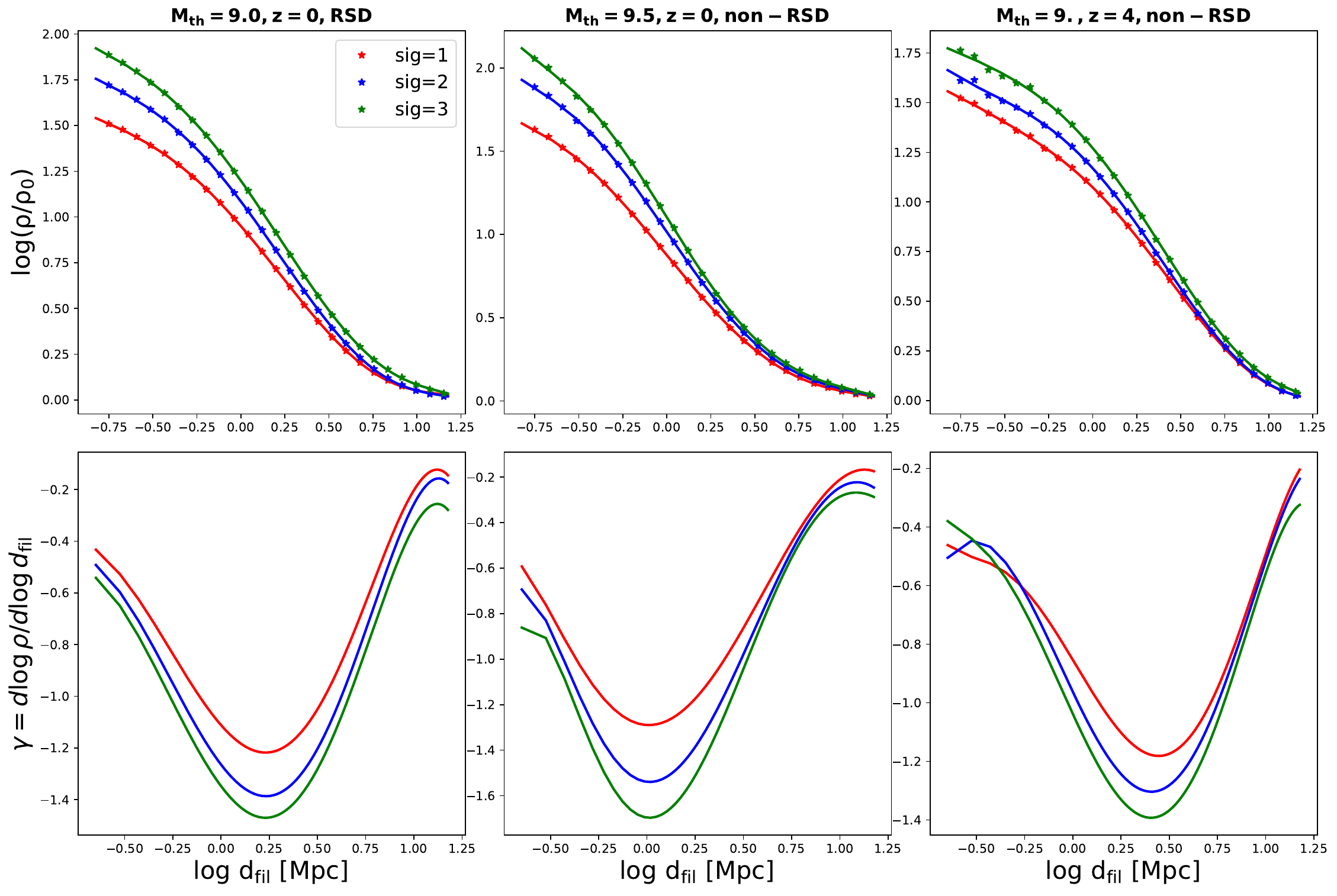}}
\caption{The profiles and slopes for filaments extracted using different significance levels of $1\sigma$ (red lines), $2\sigma$ (blue lines), and $3\sigma$ (green lines). The results are presented for the RSD sample at $z=0$ with $M_{\rm th}=10^9\msun$ (left panels), non-RSD sample at $z=0$ with $M_{\rm th}=10^{9.5}\msun$ (middle panel) and non-RSD sample at $z=4$ with $M_{\rm th}=10^9\msun$ (right panel), respectively. }
\label{fig:para}
\end{figure*}

For fair comparisons of filament radii in different samples, we adopt the same \texttt{DisPerSE} parameters to extract the filaments in this paper for different redshifts, stellar mass thresholds, RSD and non-RSD samples. However, it is still important to investigate the effects of different \texttt{DisPerSE} parameters on our results. The main parameter is the persistence significance level, which we adopted the fiducial value of $2\sigma$. In Figure~\ref{fig:para}, we show the results of the profiles and slopes with filaments extracted with significance levels of $1\sigma$ (red lines), $2\sigma$ (blue lines) and $3\sigma$ (green lines). We investigate the cases for the RSD sample at $z=0$ with $M_{\rm th}=10^9\msun$ (left panels), non-RSD sample at $z=0$ with $M_{\rm th}=10^{9.5}\msun$ (middle panel) and non-RSD sample at $z=4$ with $M_{\rm th}=10^9\msun$ (right panel), respectively.

The galaxy number density profiles apparently depend on the choice of the DisPerSE parameters with higher amplitudes and steeper slopes for filaments extracted with higher significance levels. It is therefore essential to compare the profiles of the filaments extracted with the same DisPerSE parameters in different studies. However, the positions of the slope minimum (i.e. the filament radius) are not significantly affected by the different parameter choices.

\begin{figure*}
\centerline{\includegraphics[width=0.8\textwidth]{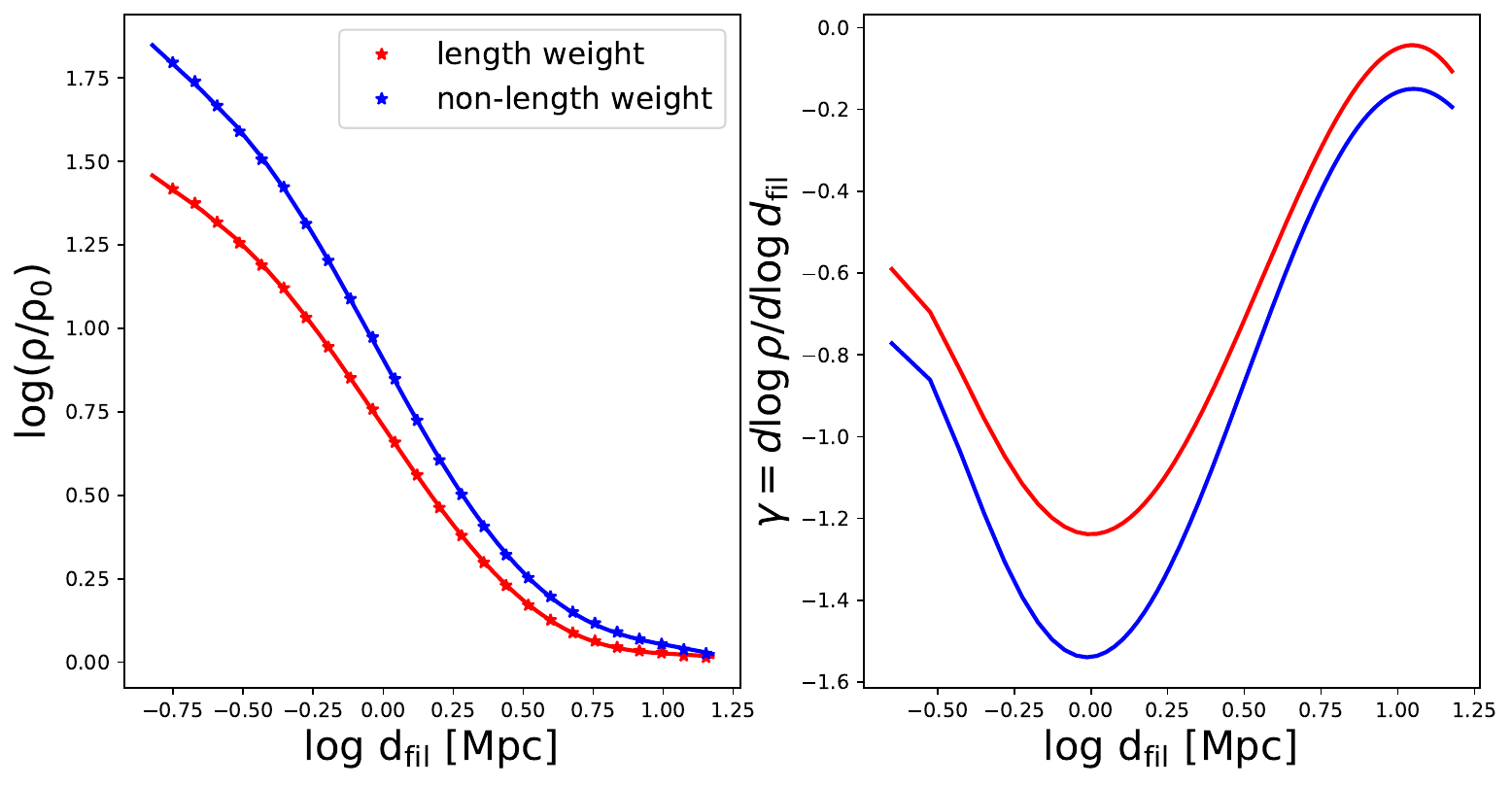}}
\caption{The profiles and the corresponding slopes stacked with (red lines) and without (blue lines) weighting by the segment length (see text for details). }
\label{fig:stack}
\end{figure*}
Another issue that will affect our results is the method to stack the filaments. We adopt the same stacking method as in \cite{2020A&A...641A.173G} and \cite{Galarraga2024} by directly averaging galaxy number density profiles for filaments of different lengths. The other potential way to stack the galaxy distributions around the filaments is to calculate the average profiles weighted by the segment lengths. We implement both methods for the \texttt{MTNG (non-RSD)} sample at $z=0$ shown in Figure~\ref{fig:stack}. The results with and without weighting by the length are shown as the red and blue lines, respectively. Our fiducial stacking method without the weight of segment length results in a higher number density profile, as the average profile is contributed more by the shorter filament segments. But the resulting filament radii in the two methods are still consistent with each other, further confirming that our results are not affected by the adopted stacking method. 

\end{document}